%% file: a0pi.tex
\documentclass[11pt]{article}
\usepackage{graphicx}

\newcommand{\BaBarYear}       {01}
\newcommand{\BaBarNumber}     {07}
\newcommand{\SLACPubNumber} {8930}

 \newcommand{\BaBarType}     {CONF}  

\input ourdefinitions

\input pubboard/babarsym

\long\def\inst#1{\par\nobreak\kern 4pt\nobreak
    {\it #1}\par\vskip 10pt plus 3pt minus 3pt}

\begin{document}
{\pagestyle{empty}

\begin{flushright}
\babar-\BaBarType-\BaBarYear/\BaBarNumber \\
SLAC-PUB-\SLACPubNumber \\
hep-ex/0107075 \\
July, 2001
\end{flushright}

\par\vskip 3cm

\begin{center}
{\Large \bf \boldmath
Search for $B^0\to a_0^+(980)\pi^-$}
\end{center}
\bigskip

\begin{center}
\large The \babar\ Collaboration\\
\mbox{ }\\
July 26, 2001
\end{center}
\bigskip \bigskip

\begin{center}
{\large \bf Abstract}
\end{center}
We present preliminary results of a search for the decay
$B^0\to a_0^+(980)\pi^- $ among 22.7 million \upsbb\ pairs 
collected by the \babar\ detector at \pep2.
Using improved background suppression techniques and optimal
signal extraction for rare decay searches, an excess of events 
over expected background is observed at the level of 3.7 
standard deviations. This corresponds to the branching fraction 
${\BR}(B^0\to a_0^+ (a_0^+ \rightarrow \eta \pi^+) \pi^-)
=(6.2_{\,-2.5}^{\,+3.0}\pm 1.1)\times 10^{-6}$, where the 
first error is statistical and the second is systematic.
The $90\%$ confidence level upper limit is $11.5\times 10^{-6}$.
\vfill
\begin{center}
Submitted to the\\ $20^{th}$ International Symposium on 
Lepton and Photon Interactions at High Energies \\
7/23/2001---7/28/2001, Rome, Italy
\end{center}
\vspace{1.0cm}
\begin{center}
{\em Stanford Linear Accelerator Center, Stanford University,
Stanford, CA 94309} \\ \vspace{0.1cm}\hrule\vspace{0.1cm}
Work supported in part by Department of Energy contract DE-AC03-76SF00515.
\end{center}
\newpage

}
\input{pubboard/authors_EPS2001}

\setcounter{footnote}{0}

%
\input{introduction}
\input{detector}
\input{method}
\input{cutcount}

\input{maxlik}
\input{systematics}

\input{conclusions}

\subsection*{Acknowledgements}
\input pubboard/acknowledgements

\input{cutopt}

\input{bibliography}
\end{document}

%% file: ourdefinitions.tex
\newcommand{\bit}{\begin{itemize}}
\newcommand{\eit}{\end{itemize}}
\newcommand{\beq}{\begin{equation}}
\newcommand{\eeq}{\end{equation}}
\newcommand{\beqn}{\begin{eqnarray}}
\newcommand{\eeqn}{\end{eqnarray}}
\newcommand{\beqns}{\begin{eqnarray*}}
\newcommand{\eeqns}{\end{eqnarray*}}

\newcommand{\intl}{\int\limits}

\def\rs{\raisebox{1.5ex}[-1.5ex]}

\def\babar{$\mbox{\sl B\hspace{-0.4em} {\small\sl A}\hspace{-0.37em} \sl B\hspace{-0.4em} {\small\sl A\hspace{-0.02em}R}}$}

\def\BB{$B\overline{B}$}

\def\BR{{\cal BR}}

\def\xout{x_{\rm out}}
\def\xcut{x_{\rm out}^{\rm cut}}
\def\xopt{x_{\rm out}^{\rm opt}}
\def\CL{{\rm CL}}
\def\CLcut{\CL^{\rm cut}}

\def\ie{{\em i.e.}} 
\def\cf{{\em c.f.}} 
\def\eg{{\em e.g.}}

\def\Nev{{\cal N}}

\def\fss{{f}}

\def\Ns{N_s}
\def\Nb{N_b}

\def\Lshape{{\cal L}_{\fss}}

\def\pss{p_s}
\def\pbb{p_b}

%% file: pubboard/authors_EPS2001.tex
\begin{center}
\small

The \babar\ Collaboration,
\bigskip

B.~Aubert,
D.~Boutigny,
J.-M.~Gaillard,
A.~Hicheur,
Y.~Karyotakis,
J.~P.~Lees,
P.~Robbe,
V.~Tisserand
\inst{Laboratoire de Physique des Particules, F-74941 Annecy-le-Vieux, France }
A.~Palano
\inst{Universit\`a di Bari, Dipartimento di Fisica and INFN, I-70126 Bari, Italy }
G.~P.~Chen,
J.~C.~Chen,
N.~D.~Qi,
G.~Rong,
P.~Wang,
Y.~S.~Zhu
\inst{Institute of High Energy Physics, Beijing 100039, China }
% ...

\end{center}\newpage

%% file: introduction.tex
\section{Introduction}
\label{sec:Introduction}

We present a search for the decay $B^0\to a_0^+(980)\pi^-$,
where the $a_0$ resonance\footnote
{
        Charge conjugation is implied throughout this document,
        and the $a_0^+(980)$ resonance is denoted  $a_0$.
} 
is observed in its dominant decay channel $a_0^+ \to \eta \pi^+$. 
The interest in this mode stems from its 
potential use for measuring the CKM angle 
$\alpha$~\cite{ref:DigheKim,ref:SardinneVasia}. 
It was pointed out in Ref.~\cite{ref:SardinneVasia}
that, within the factorization assumption, the main tree contributions 
to the $a_0\pi$ decay amplitude vanish: they would imply forbidden second  
class currents. This simplifies the two-body analysis for the extraction
of $\alpha$, and can lead to an enhanced direct \CP violation.
%

After a preselection, we refine
the background suppression using Multivariate Analyzer (MVA)
tools (Neural Net (NN) and Fisher discriminants).  
We use both cut-based and shape-based analyses, the latter employing
a maximum likelihood technique, which together provide complementary 
results and cross-checks. The cut optimization procedure 
used in the analysis does not rely on a prior branching fraction estimate.
Both inclusive and exclusive control samples are used for 
systematics checks, particularly with regard to the $\eta$ and 
$a_0$ resonances~\cite{ref:pdg2000}.

%% file: detector.tex
\section{The \babar\ Detector and the Dataset}
\label{sec:babar}

The data used in this analysis were collected with the \babar\ detector
at the \pep2\ storage ring, located at the Stanford Linear Accelerator 
Center (SLAC). \pep2\ is an asymmetric \epem\ collider with a 
center-of-mass energy equal to the \FourS\ mass.
From Nov. 1999 to Oct. 2000, a total of 22.7 
million \BB\ pairs has been collected by \babar, corresponding to 
an integrated on-peak luminosity of approximately $20.7$\invfb.
In addition, $2.6$\invfb\ of off-peak data 
were taken during the same period: they have been used  
to validate the contribution to backgrounds resulting from \epem\ 
annihilation into light \qqbar\ pairs.
 
The \babar\ detector and its performance are described in
Ref.~\cite{ref:babar}. The innermost component consists of a 5-layer 
Silicon Vertex Tracker (SVT), providing the positions of charged 
tracks in the neighbourhood of the beam interaction point. It is 
followed by a 40-layer central Drift Chamber (DCH), immersed 
in a $1.5$-T magnetic field, measuring the track momenta and 
providing a measurement of the specific ionization loss (\dedx) 
for particle identification (PID). The main PID device
is a unique, internally reflecting ring imaging Cherenkov detector 
(DIRC), covering the central region of \babar. A Cherenkov angle
$K/\pi$ separation of better than 4 standard deviations is achieved
for tracks below 3\gevc\ momentum. Photons are detected by
a CsI(Tl) electromagnetic calorimeter (EMC), which provides excellent
angular and energy resolution with high efficiency for energy deposits
above 20\mev. A superconducting solenoid, located around the EMC
is itself surrounded by an iron flux return, instrumented with 
Resistive Plate Chambers to identify muons. 

%% file: method.tex
\section{Analysis Method}
\label{sec:Analysis}

\subsection{Candidate Selection}\label{sec:preselection}

A $B^0\to a_0^+\pi^-$ candidate contains a pair 
of oppositely-charged pions and two photons.
Charged tracks are required to satisfy a set of track quality
criteria, which includes cuts on their momenta (less than 10\gevc), 
transverse momenta (above 100\mevc), and on the number of DCH hits 
(at least $20$). The tracks are also required to originate in the
vicinity of the beam-beam interaction point. 
Pion candidates must fail electron selection criteria
and are required to have a momentum in the center-of-mass (CM) 
above 2\gevc. A $B$ candidate is rejected if the track not 
used to form the $a_0$ has a
DIRC Cherenkov angle consistent with a kaon.
Photons are identified as energy deposits in the EMC, unassociated 
with charged tracks. They are required to have an energy above 
80\mev\ in the laboratory frame (LAB), and must satisfy 
photon-like shower profile criteria. To be associated with an 
$\eta$ decay, a pair of candidate photons is required to have 
an invariant mass $0.470 <m(\gamma\gamma)<0.615$\gevcc,
and the $\eta$ CM momentum must be larger 
than 0.9\gevc. The pion track and $\eta$ candidate form an $a_0$ 
candidate if their invariant mass falls in the range
$0.9 <m(\eta \pi)<1.08$\gevcc. Reconstruction of  \Bz\ candidates 
is done by vertexing all combinations of $\pi^+\pi^-\eta$ 
candidates in each event and applying a quality requirement 
on the $\pi^+\pi^-$ vertex. A \Bz\ candidate
is characterized by two kinematic variables: the beam
energy-constrained mass $m_{\rm EC}=\sqrt{E_{\rm beam}^2-p_B^2}$, where
$E_{\rm beam}$ is half the CM energy
and $p_B$ is obtained by applying kinematic constraints to the four-momenta
of the $B$ daughters; and $|\Delta E|\equiv |E_B-E_{\rm beam}|$ 
with $E_B$ being the CM energy of the $B$ candidate. A candidate 
is retained if $m_{\rm EC}>5.21$\gevcc\ and $|\Delta E|<0.25$\gev.

\subsection{Background Suppression}

Charmless hadronic modes suffer from large amounts of background 
from random combinations of tracks, mostly from light quark 
production. In the CM frame, this background 
typically exhibits a two-jet structure in contrast to the 
spherically symmetric \upsbb\ events.  Efficient background 
rejection is obtained by requiring the angle $\theta_{\rm T}$ between 
the thrust axis of the $B$ candidate and the thrust axis of the 
rest of the event (ROE) to satisfy $|\cos\theta_{\rm T}|<0.9$. 
Denoting $|\cos\theta_{\rm TP}|$ as the minimum cosine of 
the two angles formed by the two most energetic tracks (or neutral 
clusters) with respect to the thrust axis of the event, we 
require $|\cos\theta_{\rm TP}|<0.88$.
Combinatorial background within a candidate event arises mainly 
from low-energy photons. Compared with decay modes
containing $\pi^0$'s, this is a minor concern for $a_0\pi$
due to the higher $\eta$ mass: the fraction of events with more 
than one photon pair combination passing the selection cuts 
is at the percent level. 
For events with multiple candidates the one with the most 
energetic low-energy photon is retained.

\begin{figure}[t]
\begin{center}
\includegraphics[height=8cm]{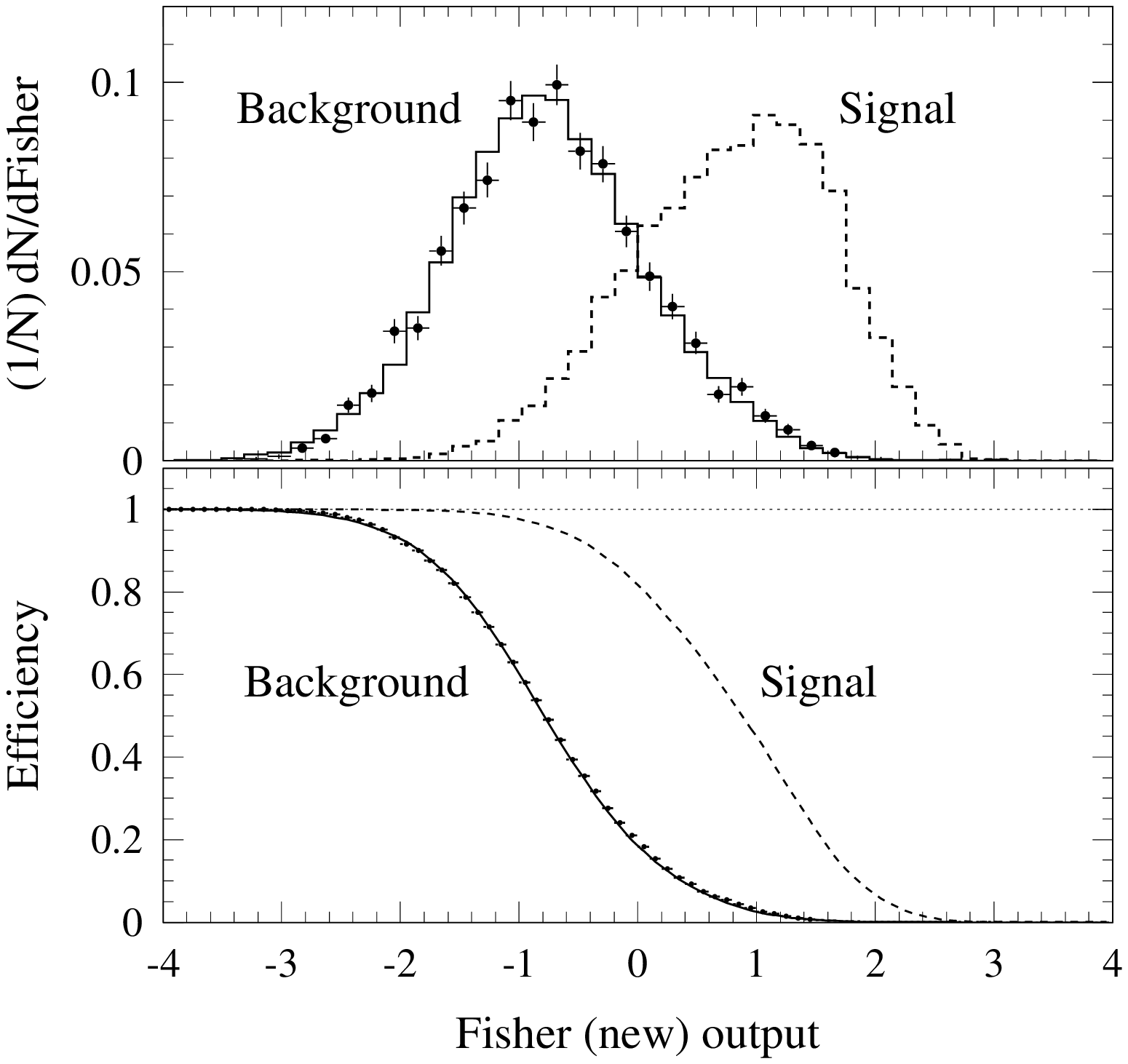}
\includegraphics[height=8cm]{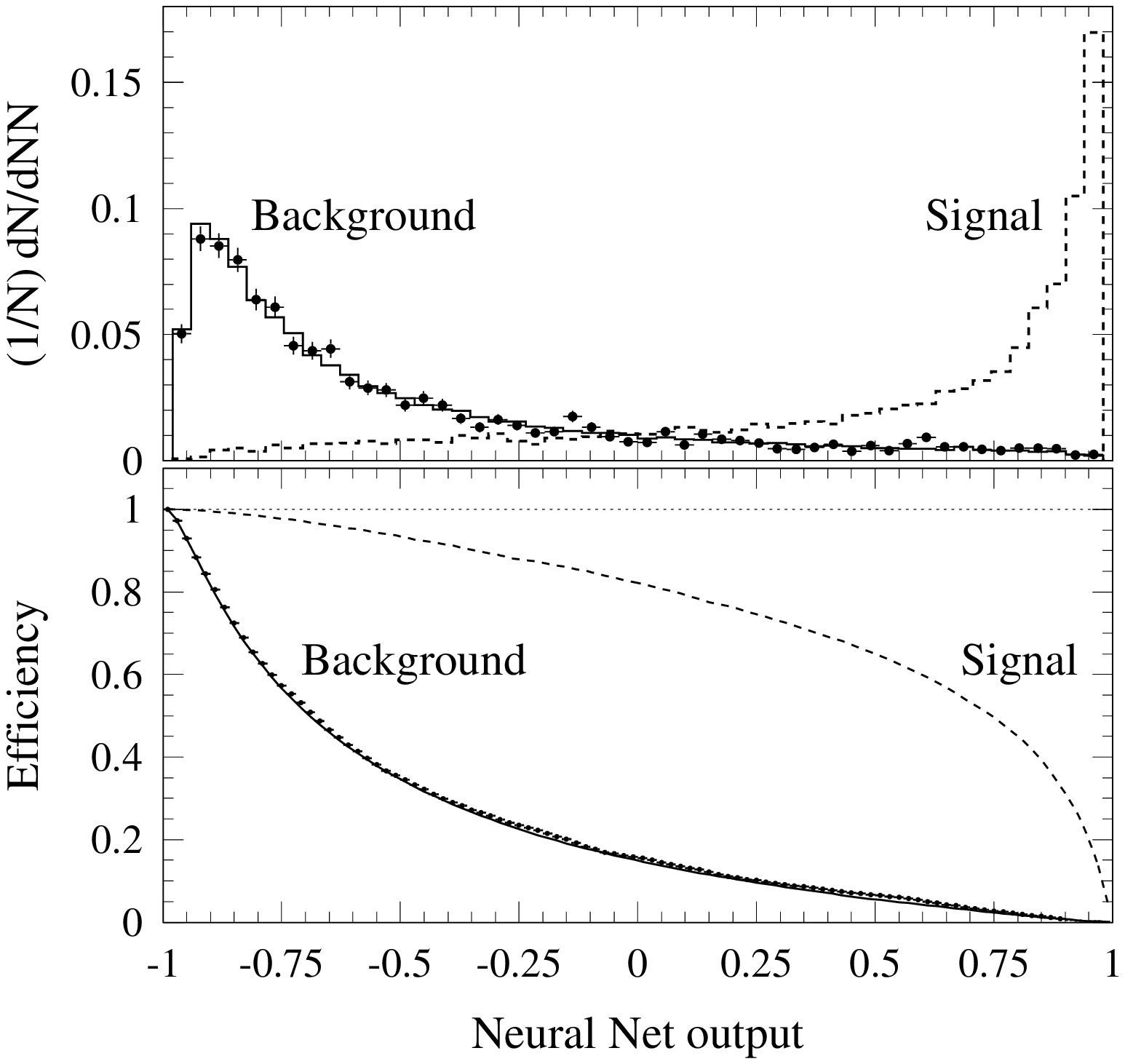}
\caption{Fisher (left-hand plots) and Neural Net discriminants
	(right-hand plots). The upper plots show the distributions
	for signal and background, where the signal is taken
        from Monte Carlo simulation and the background from on-peak 
	sideband data. The points with error bars show the off-peak 
	data, containing no signal. The lower plots show the 
	signal and background efficiencies as a function of the 
	cut applied (the small dots indicate the off-peak data).}
\label{fig:compboth}
\end{center}
\end{figure}
Further discrimination is achieved by combining event shape 
variables using MVA techniques. 
A common approach uses the Fisher discriminant (denoted 
{\em standard Fisher} below) proposed by the CLEO 
Collaboration~\cite{ref:CLEOFisher}. This standard Fisher is 
built as a linear combination of the cosine of the angle between 
the $B$ candidate momentum and the beam axis, the cosine of the 
angle between the $B$ candidate thrust axis and the beam axis,
together with nine energy deposits, each defined to be the energy 
of charged tracks and neutral clusters of the ROE
whose directions are contained in nine concentric cones
centered around the $B$ direction.
In our analysis, the choice of variables has been reconsidered 
taking into account the separation power\footnote
{
	The separation power of the normalized signal and background 
	distributions, $S(x)$ and $B(x)$, of a discriminating
	variable $x$, is defined by
	\beqns
        \langle s^2\rangle=\frac{1}{2}\intl_{-\infty}^{+\infty}\!dx
        \frac{(S(x)-B(x))^2}{S(x)+B(x)}~.
	\eeqns
}, 
the correlations between the variables and the signal efficiency 
at fixed background rejection. For this purpose,
12 variables are selected and combined using either 
a linear (Fisher) or a non-linear (NN) MVA.
Table~\ref{tab:12variables} lists the variables which are
retained together with their respective Fisher coefficients.
Among thse are the six variables $L_j^{({\rm c})}$,
$L_j^{({\rm n})}$ ($j=0,2,6$) defined by
\beq
	L_j^{({\rm c,n})}=\sum_{i_{({\rm c,n})}}p_i \times |\cos\theta_i|^j~,
\eeq
which are the momentum-weighted sums of the cosines of the angles 
between the ROE charged tracks ($L_j^{({\rm c})}$) or neutral clusters
($L_j^{({\rm n})}$) and the thrust axis 
of the $B$ candidate. These variables provide a generalization 
of the discrete cones used in the standard Fisher discriminant.

\begin{figure}[t]
\begin{center}
\includegraphics[height=8cm]{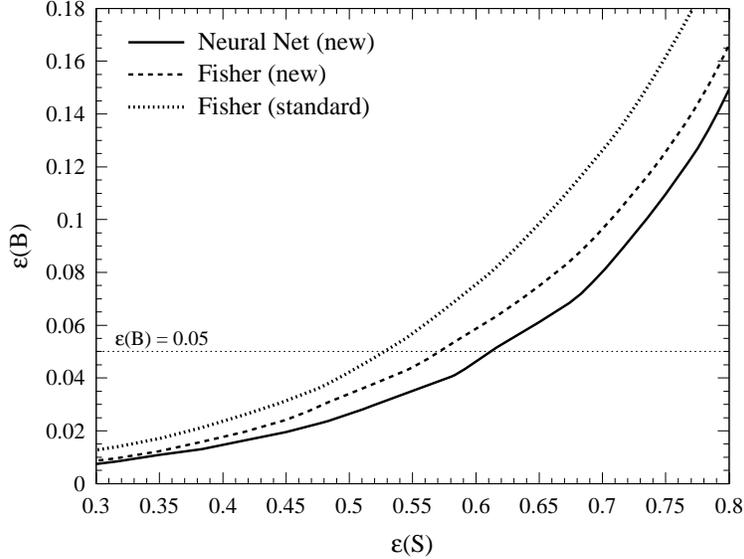}
\caption[.]{Comparison of the background versus signal efficiencies 
	for NN (solid line), the Fisher (dashed line), 
	and the standard Fisher	(dashed-dotted line) discriminants.}
\label{fig:effbs_new}
\end{center}
\end{figure}
Figure~\ref{fig:compboth} shows the distributions of the Fisher and 
NN discriminants for signal and background, the former taken from
Monte Carlo simulation and the latter from on-peak sideband data 
(histograms) and from off-peak data (points with error bars).
The lower plots show the  resulting signal and background 
efficiencies as a function of a cut applied on the output values 
of the discriminants. Figure~\ref{fig:effbs_new} depicts the 
background versus signal efficiencies for the standard Fisher 
discriminant and the 12-variable MVAs adopted 
in this analysis. A 17\% relative increase of the signal 
efficiency is obtained for NN with respect to the standard 
Fisher at the benchmark of 5\% background retention.   
Table~\ref{tab:MVAperf} summarizes the performances of the three 
MVA discriminants. The NN, being the discriminant which provides the
best signal efficiency, is used for the signal extraction in this 
analysis, while keeping the Fisher for cross-checks.
\begin{table}[t]
\caption[.]{The 12 event-shape variables used as MVA inputs
	with their corresponding Fisher coefficients.\\}
\begin{center}
\begin{tabular}{llc}
\hline
Variable name & Description & Fisher coefficient
\\
\hline
$R_2$ & Second Fox-Wolfram moment & $0.40$
\\
$|\cos\theta_{\rm T}|$& Angle: $B$/ROE thrust axis & $-0.71$
\\
$|\cos\theta_{\rm S}|$ & Angle: $B$/ROE sphericity axis& $-0.89$
\\
$|\cos(B,z)|$ & Angle: $B$ direction/beam axis & $-0.88$
\\
$|\cos(B({\rm T}),z )|$ & Angle: $B$ thrust/beam axis & $-0.79$
\\
$|\cos\theta_{\rm TP}|$ & Minimum cosine of ROE tracks/clusters &$0.54$
\\
$L^{\rm n}_0$ & Neutral zeroth-order angular function & $0.35$
\\
$L^{\rm n}_2$ & Neutral second-order angular function & $-1.02$
\\
$L^{\rm n}_6$ & Neutral sixth-order angular function& $-0.69$
\\
$L^{\rm c}_0$  & Charged zeroth-order angular function & $0.38$
\\
$L^{\rm c}_2$ & Charged second-order angular function & $-0.51$
\\
$L^{\rm c}_6$ & Charged sixth order angular function & $-0.66$
\\ \hline
Offset & {\em Centers the sum of signal and background at zero} & $1.49$
\\ \hline
\end{tabular}
\end{center}
\label{tab:12variables}
\end{table}
\begin{table}[ht]
\caption[.]{Results on the separation and the signal efficiency 
	at the benchmark of 5\% background retention for the 
	standard Fisher discriminant and the two MVAs used in
	this analysis.\\ }
\begin{center}
\begin{tabular}{lcc}
\hline
Type      	        &   $\langle s^2\rangle$
	& $\epsilon(S)({\rm at}~\epsilon(B)=5\%)$ \\ \hline
Fisher (standard) 	&   $0.44$ & $0.52$ \\
Fisher (12-var.)    	&   $0.48$ & $0.57$ \\
Neural Net (12-var.)   	&   $0.50$ & $0.61$ \\
\hline
\end{tabular}
\end{center}
\label{tab:MVAperf}
\end{table}

%% file: cutcount.tex
\subsection{Cut-Based Analysis}
\label{sec:cutandcount}

The signal extraction is based on counting data events 
found in a signal region (SR) and subtracting from this the 
expected background yield. To find an optimal set of signal 
selection criteria, balancing between background rejection
and signal efficiency, we use the procedure described in 
the appendix. The criterion it applies is to
minimize the expected confidence level for the null hypothesis 
``there is no signal in the data sample''. Cut optimization is
performed on the MVA output only, resulting in the requirements
$x_{\rm opt}^{\rm out}({\rm NN})>0.6$ and 
$x_{\rm opt}^{\rm out}({\rm Fisher})>0.7$, respectively.

\begin{table}[t]
\caption[.]{\label{tab:FinalCC} Selection efficiencies, expected
	signal and background yields for the cut-based analysis
	using Fisher or NN. When two errors are given, the
	first is statistical and the second is systematic.
	We quote upper limits as the main results
	and compute branching fractions and statistical
	significances, should  the observed 
	excess be interpreted as evidence for a signal. 
	The systematic errors are discussed in 
	Sec.~\ref{sec:systematics}.\\}
\begin{center}
\begin{tabular}{lcc}
\hline
 & Fisher & Neural Net        \\ \hline
Signal efficiency ($\%$) & $14.2$ & $14.6$  \\ 
Events in signal box & 20 & 18            \\ 
Events in GSB & 242 & 197      \\ 
Expected background  & & \\
events in signal box 
	& \rs{$9.6\pm0.6\pm1.4$} 
	& \rs{$7.9\pm0.6\pm1.1$} \\ 
Branching fraction $(\times 10^{-6})$
	& $8.2_{\,-3.0}^{\,+4.6}\pm1.4$ 
	& $7.8_{\,-2.7}^{\,+4.0}\pm1.2$ \\
Statistical significance & $3.0\sigma$ 	& $3.1\sigma$ \\ 
$90\%$ CL Upper limit $(\times 10^{-6})
	$ & $15.3$ & $14.5$ \\
\hline
\end{tabular}
\end{center}
\end{table}
The SR is defined as $|m_{\rm EC}-5.28|<0.006$\gevcc, 
$|\Delta E|<0.070$\gev\ and $0.517<m(\gamma\gamma)<0.587$\gevcc.  
The background contamination in the SR is estimated from the 
two-dimensional Grand Sideband (GSB): $|\Delta E|<0.25$\gev\ and 
$5.210<m_{\rm EC}<5.263$\gevcc, assuming for the corresponding 
background shapes a second-order polynomial for $\Delta E$, 
and an ARGUS function~\cite{ref:argusshape} for $m_{\rm EC}$. 
The signal efficiencies ($\epsilon$), event yields ($\Nev$) 
and the expected backgrounds ($\Nb$) are given in 
Table~\ref{tab:FinalCC} for Fisher and NN. The corresponding
branching fractions are obtained from the expression
\beq
\label{eq:brcc}
	{\BR} = \frac{\Nev - \Nb}{\epsilon\, N_{B\bar B}}~,
\eeq
where $N_{B\bar B}=(22.74\pm 0.36)\times 10^6$ is the total
number of produced \BB\ pairs. Equation~\ref{eq:brcc} assumes 
an equal production of charged and neutral $B$'s in \FourS\ decays. 
Non-resonant 
contributions to the $\eta\pi^+\pi^-$ final state are 
disregarded for all branching fractions quoted in this document.
The results obtained using Fisher and NN are compatible. Their 
statistical significances, defined as the probability to observe 
the given excess when no signal is present, are 3.0 and 3.1
standard deviations, respectively. The systematic errors assigned
to the branching fractions are described in Sec.~\ref{sec:systematics}.
Also given in Table~\ref{tab:FinalCC} are the upper limits
at 90\%~ confidence levels, where the systematic
errors are added linearly to the statistical limits.
\begin{figure}[!h]
\begin{center}
\vspace{-1cm}
\includegraphics[height=9cm]{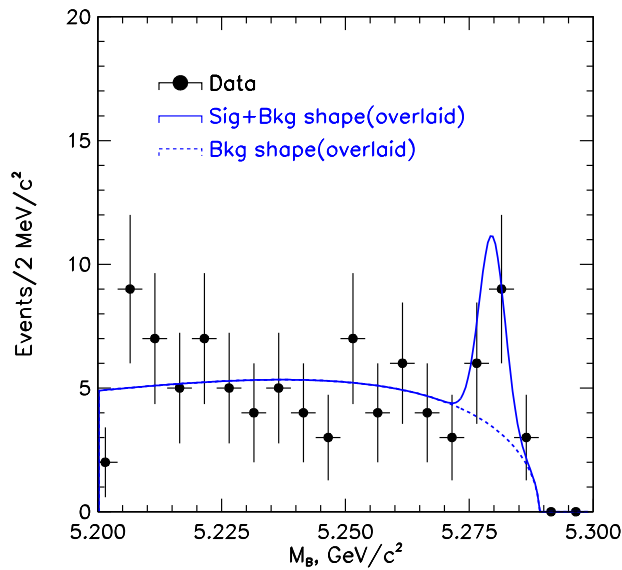}
\caption{Distribution of $m_{\rm EC}$ obtained from the cut-based analysis 
        using the NN. Superimposed on the data points are
	the signal and background $m_{\rm EC}$ distributions,
	normalized to the number of observed events. }
\label{fig:nn_overlay}
\end{center}
\end{figure}

The $m_{\rm EC}$ distribution of the events passing the selection
requirements (except for the $m_{\rm EC}$ cut) 
is shown in Fig.~\ref{fig:nn_overlay} (for NN). For 
illustration purposes, we have superimposed the Gaussian signal 
contribution and the ARGUS background contribution;
both distributions are normalized to the results given in 
Table~\ref{tab:FinalCC}.

%% file: maxlik.tex
\subsection{Maximum Likelihood Analysis}
\label{sec:maximumLikelihoodAnalysis}

The likelihood $P(\fss,x_i)$ for event $\{i\}$ with a
measured set of discriminating variables $x_i$, and for
a signal fraction $\fss$, is defined as
\beq
   P(\fss,x_i) = \fss \pss(x_i) + (1-\fss) \pbb(x_i)~.
\eeq
The $\pss(x_i)$ ($\pbb(x_i)$) is the product of the normalized 
signal (background) probability density functions (PDF) 
of the individual variables entering the fit: namely,
$m_{\rm EC}$, $\Delta E$, $m(\eta)$ and the NN or Fisher output.
The $\eta\pi$ invariant mass is not included since the 
line shape of the $a_0$ resonance is not well known at 
present~\cite{ref:pdg2000}.
The data sample is selected within generous sidebands for
the variables that enter the fit (\cf\ Sec.~\ref{sec:preselection}).
\begin{figure}[t]
\begin{center}
\includegraphics[height=8cm]{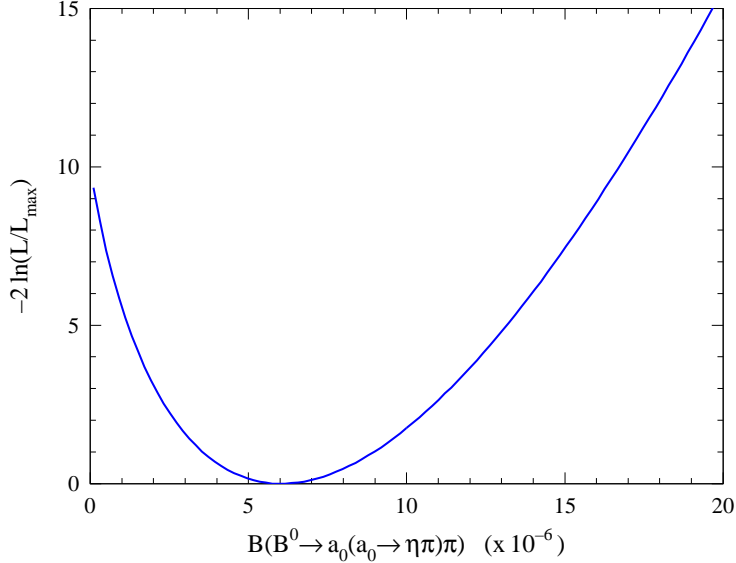}
\caption{Negative log-likelihood function versus the
	signal branching fraction.}
\label{fig:likelihood}
\end{center}
\end{figure}
The signal distributions are obtained from Monte Carlo 
simulation, refined with data from the signal-like charmed $B$ 
decay mode $D\rho$, and inclusive samples. On-peak
sideband events, controlled by off-peak data, are used to 
infer the corresponding background shapes. 
Multi-Gaussian, polynomial functions and cubic splines have been 
used to empirically approximate the reconstructed shapes of 
the discriminating variables. 

After applying the event selection described in 
Sec.~\ref{sec:preselection}, a total of $\Nev=9248$ candidate events 
without multiple combinations enters the ML fit, corresponding to a 
signal efficiency of 32.8\%. The fit is performed by minimizing 
the sum 
\beq
 -2\ln\Lshape=-2\sum_{i=1}^{\Nev} \ln P(\fss,x_i)~,
\eeq
over $\Nev$ events, with respect to the signal fraction $\fss$. 
The fit results in 
$\Ns=\fss\Nev=18.1_{\,-7.4}^{\,+8.7}$ signal events for NN, and 
$\Ns=16.2_{\,-7.2}^{\,+8.6}$ signal events for Fisher\footnote
{
	From ``toy'' Monte Carlo studies, the difference 
	of 1.9 events observed between the signal yields with 
	NN and Fisher is consistent with the statistical overlap 
        between the samples.
	Taking into account the 84\% correlation
	between the signal yields of the fits with NN and Fisher,
	the probability to find a difference larger than that observed 
	is 45\%.
}. 
In the following, we restrict the discussion to the results obtained with the 
NN, since it provides the best signal 
efficiency. This choice has been made prior to uncovering the 
results. The negative log-likelihood function 
versus the signal branching fraction is depicted in
Fig.~\ref{fig:likelihood}. A ``toy'' Monte Carlo simulation 
indicates a goodness-of-fit of 50\%.
If the observed signal excess is interpreted as evidence for a 
signal, the corresponding branching fraction is
\beq
\label{eq:ml_br}
        {\BR}(B^0\to a_0^\pm
                (a_0^\pm \rightarrow \eta \pi^\pm) \pi^\mp)
                \,=\,(6.2_{\,-2.5}^{\,+3.0}
                 \pm1.1)\times 10^{-6}~,
\eeq
where the first error quoted is statistical and the second is systematic.
The statistical significance of this result corresponds to $3.7$ 
standard deviations. The latter is computed using a zero-signal 
toy Monte Carlo simulation. Systematic effects are discussed 
in Sec.~\ref{sec:systematics}. Assuming no evidence for a signal, 
the $90\%$~CL upper limit on the branching fraction, obtained from 
the integral $\int_0^{\BR} \Lshape({\BR}^\prime) d{\BR}^\prime/
\int_0^\infty \Lshape({\BR}^\prime) d{\BR}^\prime=0.9$, is
\beq
	{\BR}(B^0\to a_0^\pm 
		(a_0^\pm \rightarrow \eta \pi^\pm) \pi^\mp) 
        < 11.5 \times 10^{-6} ~,
\eeq
where the systematic uncertainty has been added linearly.
A 65\% correlation between the ML fit result~(\ref{eq:ml_br})
and the cut-based result (see Table~\ref{tab:FinalCC})
is found with a toy Monte Carlo simulation. The probability 
for finding a larger difference than that observed  
($1.6\times10^{-6}$) is found to be 54\%.

\begin{figure}[t]
\begin{center}
\includegraphics[height=8.1cm]{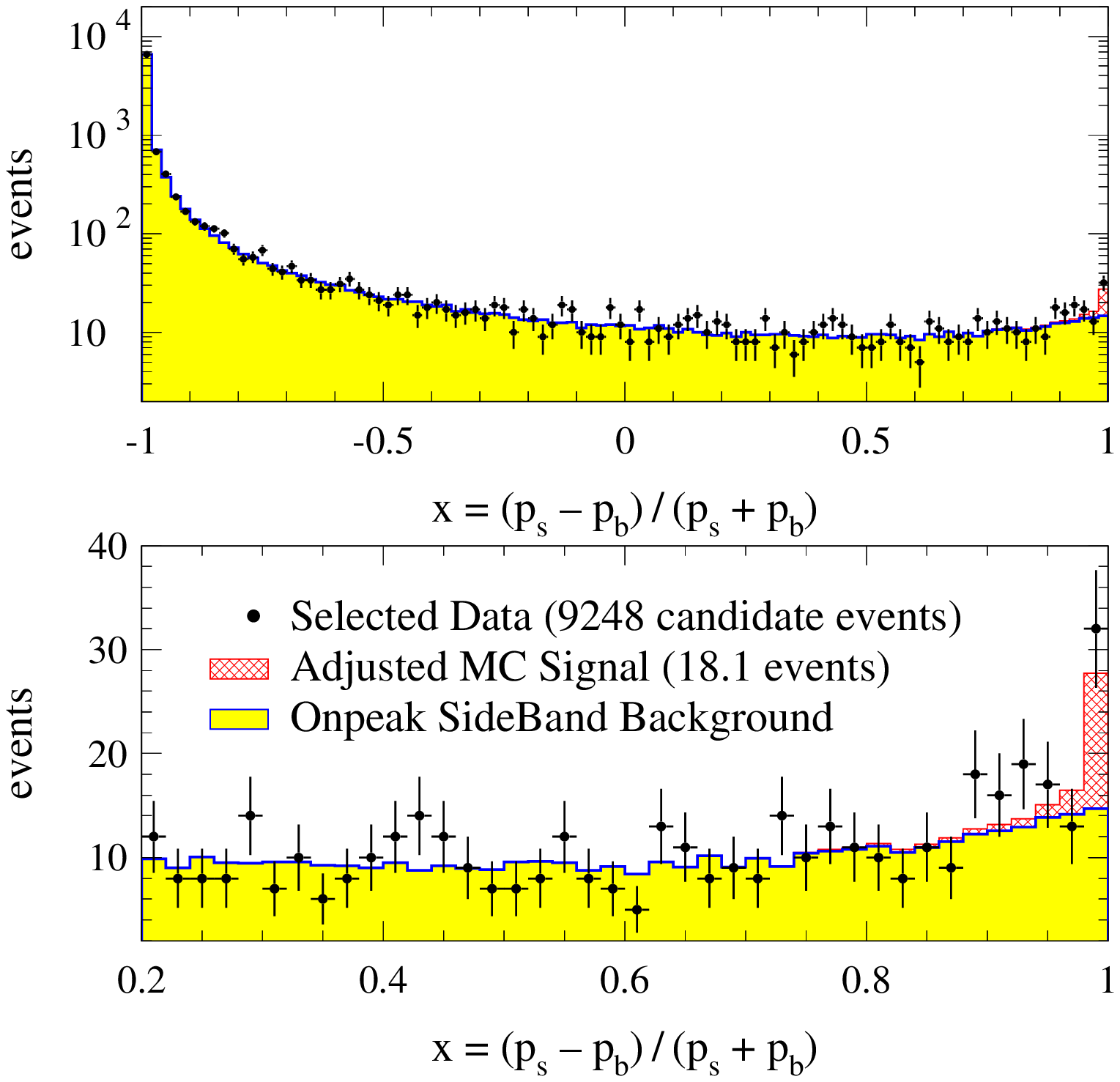}
\includegraphics[height=8.1cm]{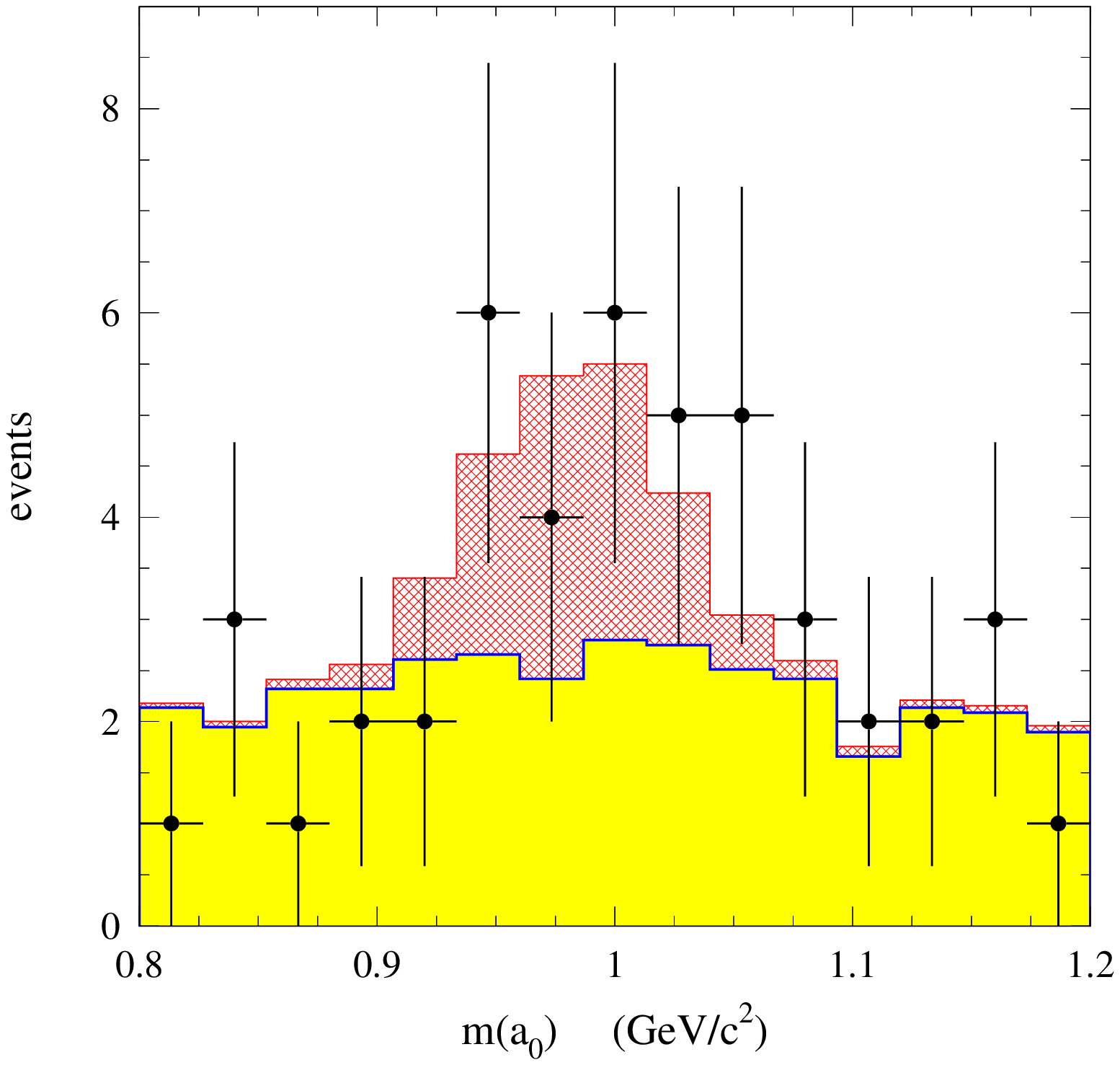}
\caption[.]{Left-hand plot:
	distributions of the $x$ variable~(\ref{eq:xvariable})
	for data (points with error bars), signal Monte Carlo 
	(cross-hatched area) and the background expectation 
	from on-peak sidebands (shaded area). 
	The signal region has been expanded in the 
	lower plot. The simulated
	signal contribution is normalized to the ML fit result.
	A signal excess is observed.
	Right-hand plot: projection on the $\eta\pi$ 
	invariant mass axis after requiring $x>0.98$, removing 
	$44\%$ of the signal and
	99.9\% of the background.}
\label{fig:xvariable_ma0proj1}
\end{center}
\end{figure}

The signal yield of the likelihood function can be 
displayed with the {\em x-variable}, defined by
\beq
\label{eq:xvariable}
	x \,= \, \frac{\pss-\pbb}{\pss+\pbb}~.
\eeq
Its distribution for the data sample used in the ML fit
is shown by the points with error bars in the left-hand plots
of Fig.~\ref{fig:xvariable_ma0proj1}.
Also shown are the expectations from signal Monte Carlo 
(cross-hatched area), normalized to the signal yield of the
ML fit, and background from on-peak sidebands 
(shaded area). A signal excess and good agreement between 
data and simulation are observed.

Since the information from the shape of the $\eta\pi$ invariant 
mass distribution is not exploited in the likelihood function, 
one can use  projections in order to verify that the 
signal candidates are consistent with the $a_0$ hypothesis. 
We select events with $x>0.98$, 
keeping $56\%$ of the signal while retaining only 0.1\%
of the background. The distributions of the corresponding
data events, as well as the signal and background expectations 
are shown in Fig.~\ref{fig:xvariable_ma0proj1}. The data are  
consistent with the expected enhancement at the $a_0$ invariant 
mass.

%% file: systematics.tex
\section{Systematic Uncertainties}
\label{sec:systematics}

The main sources of systematic uncertainty originate from the 
accuracy of the simulation for the reconstruction of neutrals, from
the tracking efficiency and from particle identification. 
Dedicated studies provide estimates 
of these effects. A $5\%$ uncertainty is assigned to the 
$\eta$ reconstruction efficiency, which represents the dominant 
error on the selection efficiency. The tracking efficiency 
difference between data and Monte Carlo simulation amounts to 
$3.9\%$. An uncertainty of  
$1.8 \%$ is assigned due to particle identification.

Other systematics may arise from the imperfect simulation of the 
distributions of the discriminating variables. They have been 
studied with control samples. The widths and central 
values of the $\Delta E$ and $m_{\rm EC}$ signal distributions 
are calibrated with $B^- \rightarrow D^0(\rightarrow K^-\pi^+) \rho^-$ 
decays, where
the estimate of systematics using final states with $\pi^0$'s
is expected to be conservative since the 
$\pi^0$'s have more 
combinatorial background than $\eta$'s. The $\eta$ and 
$a_0$ invariant mass distributions have been studied with
inclusive data samples. Since the line shape of the $a_0$ 
resonance is not well known~\cite{ref:pdg2000}, we have used 
a generous mass window (\cf\ Sec.~\ref{sec:preselection}) so 
that the resulting systematic 
effect from the $\eta\pi$ mass requirement is small. 
The Fisher and NN distributions have 
been checked by comparing on-peak sideband data and $q\overline{q}$ 
Monte Carlo, where all differences are assigned as systematic 
errors.  The uncertainty on the signal efficiency related to 
the limited Monte Carlo statistics is negligible.

The contamination from continuum background in the cut-based
analysis is evaluated by extrapolating the number of events 
found in the GSB, using the known shapes of the $\Delta E$ and 
$m_{\rm EC}$ background distributions. The consistency of the 
parameterizations has been checked for different regions of the
sidebands, and between data sidebands and $q\bar{q}$ Monte 
Carlo samples as well as off-peak data. Good agreement is found. 
Associated uncertainties are estimated by varying the shape 
parameters within their statistical accuracies. The same 
procedure has been applied to estimate the systematic 
uncertainties from the cuts on the event shape variables and 
the MVA discriminants. 

The estimate of the (mostly charmless) 
$B$-background has been performed with generic charmless Monte 
Carlo. We expect a contamination of $0.3$ events from 
non-resonant $\pi^+\pi^-\pi^0$ decays. The feedthrough from 
the unknown decay mode $B^0\rightarrow a_0^+K^-$ is 
determined from the comparison of the inclusive $a_0^+h^-$ signal
yield (\ie, obtained without PID requirements) with the exclusive 
results of the cut-based analysis. No indication for a possible 
contamination from $a_0^+K^-$ events is found.
A conservative 5\% systematic is assigned to the kaon fraction 
in the $a_0^+\pi^-$ signal.

Table~\ref{syst_tot} summarizes the  relative systematic errors 
used for the cut-based analysis. Where they differ, uncertainties 
are given separately for NN and Fisher.

Systematic effects specific to the ML analysis arise from 
differences in the background shapes of the MVA variables
between data and Monte Carlo simulation ($15\%$). The systematic 
uncertainty from the  correlations between the discriminating 
variables entering the fit, in particular those between the signal 
distributions of $\Delta E$ and $m_{\rm EC}$, amounts to $2\%$. The 
total systematic error for the ML fit is $18\%$.
\begin{table}[t]
\caption[.]{\label{syst_tot} Summary of systematic uncertainties 
	for the cut-based analysis.\\}
\begin{center}
\begin{tabular}{lc}
\hline
Source           &  Uncertainty               \\ \hline

$\eta$ finding   &         $5.0     \%$  \\ 
$\eta$ mass and resolution      
	&         $0.7  \%$   \\ 
$a_0$ mass and width/resolution       &         $0.1 \%$    \\
Track finding    &         $3.9   \%$  \\
PID              &         $2.6  \%$  \\ 
$\Delta E$       &         $1.7  \%$    \\
$m_{\rm EC}$         &         $0.1 \%$    \\ 
$\theta_{\rm T}$ and $\theta_{\rm TP}$ 
		&	$1.0\%$ \\
MVAs 
		&	NN: $1.0\%$, FI: $2.0\%$ \\
GSB $q\bar{q}$ background estimate  
		&         NN: $1.3$ events, FI: $1.5$ events  \\
Charmless $B$ background &      $0.3$ events   \\ 
$a_0^+K^-$ feedthrough 	&      $5\%$   \\ 
MC sample statistics &     $0.2  \%$  \\ 
$B$ counting      &         $1.5 \%$   \\ \hline
Total error on branching fraction            
	&  NN: $15.7 \%$, FI: $17.0 \%$ \\ \hline
\end{tabular}
\end{center}
\end{table}

%% file: conclusions.tex
\section{Conclusions}

The preliminary analysis reported in this paper shows an excess 
over expected background of $B^0\to a_0^+(980)\pi^-$ events, 
excluding the zero-signal hypothesis at the level of $3.7$ 
standard deviations. Interpreted as evidence for a signal, the 
excess would result in the branching fraction
${\BR}(B^0\to a_0^+(a_0^+ \rightarrow \eta \pi^+) \pi^-)=(6.2_{\,-2.5 }^{\,+3.0} \pm1.1)\times 10^{-6}$, where the first error
quoted is statistical and the second is systematic.
This corresponds to an upper limit of
$       {\BR}(B^0\to a_0^+
                (a_0^+ \rightarrow \eta \pi^+) \pi^-)
        < 11.5 \times 10^{-6}$ at 90\%~CL.
We emphasize the use of improved, linear and 
non-linear multivariate background suppression techniques 
and optimal signal extraction criteria for rare decay searches.

%% file: pubboard/acknowledgements.tex
We are grateful for the 
extraordinary contributions of our \pep2\ colleagues in
achieving the excellent luminosity and machine conditions
that have made this work possible.
The collaborating institutions wish to thank 
SLAC for its support and the kind hospitality extended to them. 
This work is supported by the
US Department of Energy
and National Science Foundation, the
Natural Sciences and Engineering Research Council (Canada),
Institute of High Energy Physics (China), the
Commissariat \`a l'Energie Atomique and
Institut National de Physique Nucl\'eaire et de Physique des Particules
(France), the
Bundesministerium f\"ur Bildung und Forschung
(Germany), the
Istituto Nazionale di Fisica Nucleare (Italy),
the Research Council of Norway, the
Ministry of Science and Technology of the Russian Federation, and the
Particle Physics and Astronomy Research Council (United Kingdom). 
Individuals have received support from the Swiss 
National Science Foundation, the A. P. Sloan Foundation, 
the Research Corporation,
and the Alexander von Humboldt Foundation.

%% file: cutopt.tex
\section*{Appendix: Cut Optimization}

The cut on the MVA variable output $\xout$ 
(the final cut being denoted $\xcut$) is applied following a
criterion  designed for rare decay searches. In particular, 
it does not require one to know (or guess) the branching fraction 
of the signal one is looking for.
When searching for a rare signal, one wants to rule out the  null
hypothesis\footnote
{
	It can be shown that ruling out the opposite hypothesis
	``there is  signal in the data sample'', leads to 
	an identical prescription.
}:
``there is no signal in the data sample''. For this purpose, one defines 
the confidence level
\begin{equation}
\label{eq:CL}
        \CL(N,\xcut) = \sum_{n=N}^{\infty}{\cal P}_n(N_b)~,
\end{equation}
where $N$ is the number of events  retained by the final cut,
$N_b$ is the expected number of background events passing the final cut,
and ${\cal P}_n(N_b)$ is the corresponding Poisson probability
distribution.
If $\CL$ is below a certain threshold (\eg, $5 \%$, meaning that the 
hypothesis is excluded at $95 \%$ $\CL$), one excludes the absence 
of rare decays, to this level. The idea is to adjust $\xcut$,
hence varying both signal and background efficiencies to reach the 
lowest confidence level, on average.

\paragraph{Cut Optimization for Known Branching Fraction {\BR}.}

First, if the branching fraction is known, then, for 
a given $\xcut$, one can predict the expected number of $N_s$ signal 
events. The expected value of the $\CL$ of Eq.~\ref{eq:CL} (\ie, the 
average over a large number of hypothetical experiments) is
\begin{equation}
\label{eq:CLav}
        \langle \CL\rangle(\xcut) = 
        \sum_{N=0}^{\infty} {\cal P}_N(N_s+N_b)\ \CL(N,\xcut)~.
\end{equation}
The optimal cut on $\xout$, denoted $\xopt({\BR})$, is the one which 
minimizes the above average, \ie, which leads to the clearest 
\begin{figure}[t]
\begin{center}
\includegraphics[height=6.5cm]{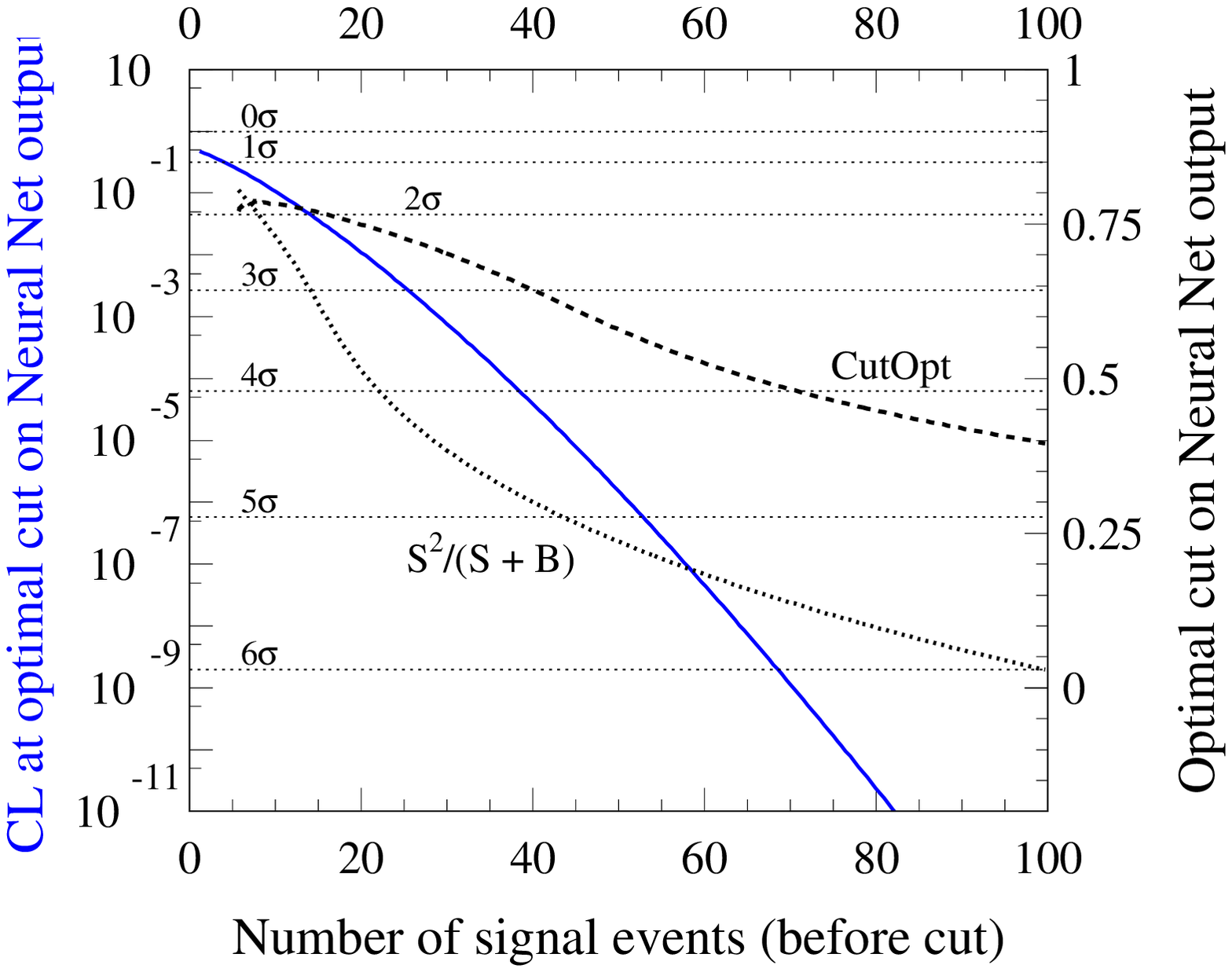}
\includegraphics[height=6.5cm]{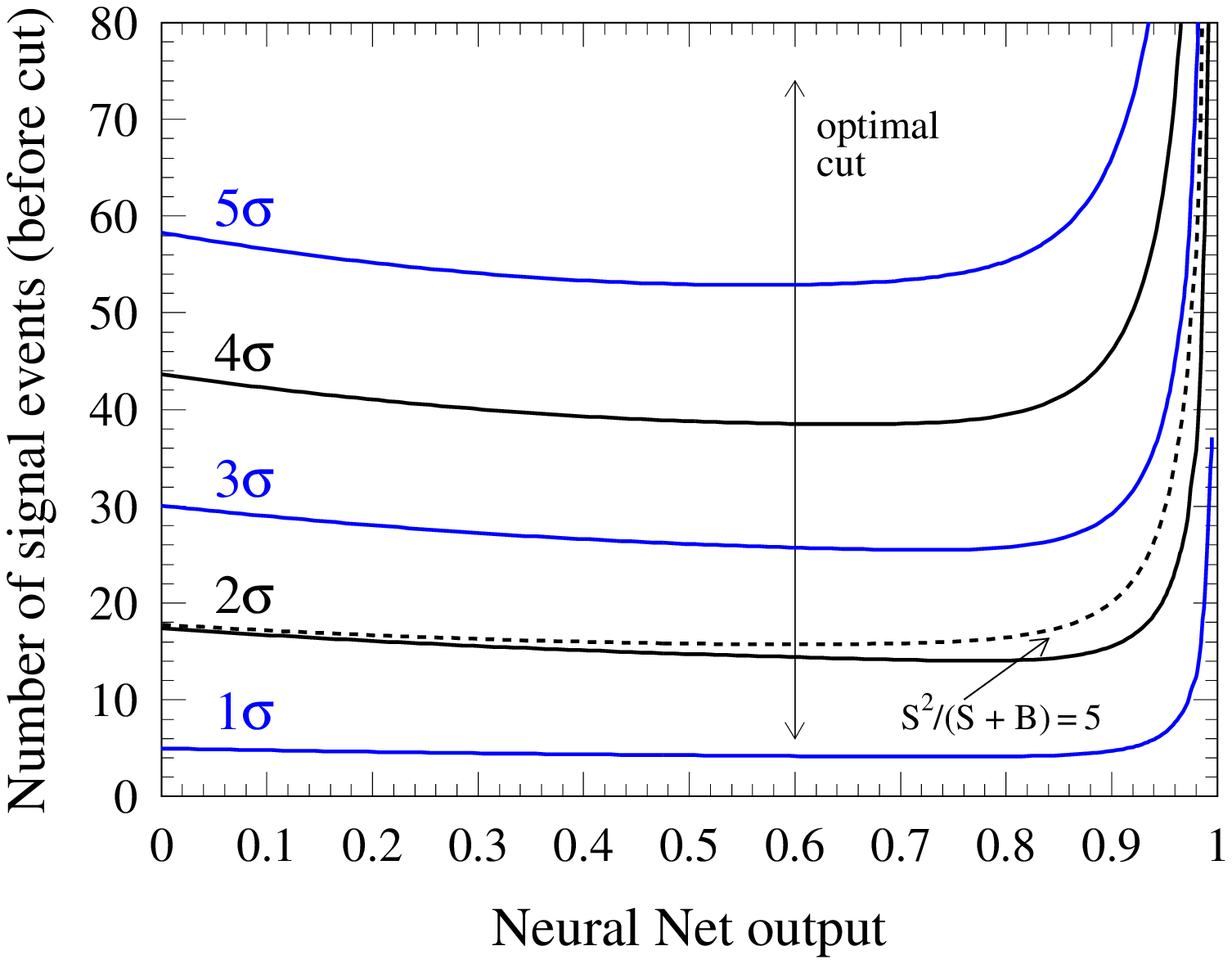}
\caption{Left-hand plot:
	expected confidence levels and optimal cuts on the NN 
	discriminant for an analysis where the branching fraction 
	is known. The solid line gives the confidence level,
        obtained when using the optimal cut, as a function of the number
        of signal events (before cut). The dashed line is the
        optimal cut for a given number of events, and the
        dotted line gives the optimal cut using the 
	criterion $S^2/(S+B)$.
	Right-hand plot:
	average number of signal events before NN cut, needed in 
	the data sample to attain a given confidence level, on
	the average. Shown are the curves corresponding
        to $1\dots 5\sigma$. The minima of the curves correspond to
        the optimal cuts at a given significance level. Notice that
        the optimal cuts are approximately independent of the
        branching fraction. Shown in addition is the curve obtained
        for the constant signal excess $S^2/(S+B)=5$,
        corresponding to a significance of 2 standard deviations.}
\label{fig:cutopt}
\end{center}
\end{figure}
rejection of the (wrong) hypothesis:``there is no signal in 
the data sample''
\begin{equation}
        \label{eq:xoptBR}
        \langle \CL\rangle(\xopt({\BR}))
        =\langle \CL\rangle_{{\rm min};\,\xcut}(\xcut)~.
\end{equation}
Figure~\ref{fig:cutopt} (left hand plot) shows the confidence level 
achievable for a given number of expected signal events. Also 
shown are the variations of the optimal cut and 
the cut obtained when using the Gaussian criterion $S^2/(S+B)$
for optimization.

\paragraph{Cut Optimization for Unknown Branching Fraction.}

When no expectation for the signal branching fraction is available,
Eq.~\ref{eq:xoptBR} cannot be used to define the optimal cut 
value of $\xcut$. The idea is then to replace the target branching 
fraction by a target  confidence level. One chooses a particular 
value for the target $\CL$ (denoted $\CLcut$) one is aiming at
to define the optimal $\xcut$ value as the value $\xopt$ 
(now without $({\BR})$ as an argument) for which the equality
\begin{equation}
\label{eq:CLcut}
	\langle \CL\rangle(\xopt) = \CLcut~,
\end{equation}
is reached for the smallest {\BR}. In the following we use 
$\CLcut=0.05$. The performance for unknown branching fraction is
illustrated for the NN discriminant in the right hand plot of 
Fig.~\ref{fig:cutopt}.

%% file: a0pi.bbl
\begin{thebibliography}{99}

\bibitem{ref:DigheKim}         A.S. Dighe, C.S. Kim , 
                        {\it Phys.Rev.D}{\bf 62}, 111302 (2000).
\bibitem{ref:SardinneVasia} 
			S. Laplace and V. Shelkov, 
                        ``{\em CP Violation and the Absence of 
                        Second Class Currents in Charmless B Decays}'', 
                        LAL 01-24, LBNL-47757, hep-ph/0105252 (2001),
			to appear in Eur. Phys. J. C.
\bibitem{ref:pdg2000}   Particle Data Group, 
			D.E.~Groom {\em et al.}, \epjc{15}, 1 (2000).
\bibitem{ref:babar}	\babar\ Collaboration,
			B.~Aubert {\em et al.}, 
			{\em ``The \babar\ Detector''},
			\hepex{0105044} (2001), to appear in \nimBaseB.
\bibitem{ref:nn}	P.~Gay, B.~Michel, J.~Proriol, and O.~Deschamps,
                        ``{\em Tagging Higgs Bosons in Hadronic LEP-2 
                        Events with Neural Networks.}'',
                        In Pisa 1995, New computing techniques in 
                        physics research, 725 (1995).
\bibitem{ref:physbook} 	\babar\ Collaboration, 
			P.F.~Harrison and H.R.~Quinn, eds.,
			{\em ``The \babar\ Physics Book''},
			SLAC-R-504 (1998).
\bibitem{ref:CLEOFisher}CLEO Collaboration, D.M.~Asner {\em et al.},
                        {\it Phys. Rev.} {\bf D53}, 1039 (1996).
\bibitem{ref:argusshape}ARGUS Collaboration, H.~Albrecht {\em et al.},
			{\em Phys. Lett.} {\bf B185}, 218 (1987);
			{\bf B241}, 278 (1990).
\end{thebibliography}
